%% file: corona-arxiv.tex
\documentclass[11pt,a4paper]{article}

\usepackage{fullpage,graphicx,latexsym,amsfonts,amssymb,amsmath,amsthm,wrapfig,subfigure}
\usepackage[hyperfootnotes=false]{hyperref}
\usepackage{authblk}

\usepackage{cite}

\input{./macros.tex}

\graphicspath{{figures/}}

\title{Infection dynamics of COVID-19 virus under lockdown and reopening}
\date{}

\author[a]{Jakub Svoboda}
\author[b]{Josef Tkadlec}
\author[c]{Andreas Pavlogiannis}
\author[a]{Krishnendu Chatterjee}
\author[b,$\star$]{Martin A. Nowak}
\affil[a]{IST Austria, A-3400 Klosterneuburg, Austria}
\affil[b]{Department of Mathematics, Department of Organismic and Evolutionary Biology, Harvard University, Cambridge, MA 02138, USA}
\affil[c]{Department of Computer Science, Aarhus University, Aabogade 34, 8200 Aarhus, Denmark}
\affil[$\star$]{To whom correspondence should be addressed. E-mail: martin\_nowak@harvard.edu}

\begin{document}
\maketitle
\begin{abstract}
Motivated by COVID-19, we develop and analyze a simple stochastic model for a disease spread in human population.
We track how the number of infected and critically ill people develops over time in order to estimate the demand that is imposed on the hospital system.
To keep this demand under control, we consider a class of simple policies for slowing down and reopening the society and we compare their efficiency in mitigating the spread of the virus from several different points of view. 
We find that in order to avoid overwhelming of the hospital system, a policy must impose a harsh lockdown or it must react swiftly (or both).
While reacting swiftly is universally beneficial, being harsh pays off only when the country is patient about reopening and when the neighboring countries coordinate their mitigation efforts.
Our work highlights the importance of acting decisively when closing down and the importance of patience and coordination between neighboring countries when reopening.

\hide{Possibly mix with: Our model can be used to advise general policies of countries and individual strategies of organizations, colleges and companies.}

\end{abstract}



\section{Introduction}
Severe acute respiratory syndrome coronavirus 2 (SARS-CoV-2) is the virus that causes the current pandemic of coronavirus disease 2019 (COVID-19).
The infection was first identified in December 2019 in Wuhan (China) and has since spread globally. 
By the end of 2020 more than 80 million people have been tested positive for the virus and more than 1.7 million people have died from complications caused by the infection.
The large majority of cases result in recovery after mild or no symptoms.
The coronavirus pandemic has led to an unprecedented global response in quarantine measurements, social distancing, travel restrictions and shutting down of economic activity. 

The use of mathematics in the study of dynamics of infectious diseases has a long tradition.~\cite{bernoulli1760essai,hamer1906epidemic,kermack1927contribution,bailey1975mathematical,diekmann1990definition,anderson1992infectious,dietz1993estimation,nowak1994superinfection,nowak1996population,diekmann2000mathematical,nowak2000virus,hethcote2000mathematics,brauer2012mathematical}
Mathematical epidemiology was immediately applied to the coronavirus pandemic~\cite{anderson2020will,ferguson2020report,fauci2020covid}, resulting in an immense amount of literature on the subject.
As a quick sample, there is ample work on understanding how the disease progresses within an individual~\cite{guan2020clinical,zhou2020clinical,grasselli2020baseline}, how it spreads through the community~\cite{li2020early,xu2020epidemiological,wu2020estimating,kucharski2020early,leung2020first,zhang2020changes,rader2020crowding}, and on the effects of varisou non-pharmaceutical interventions under varying circumstances~\cite{kraemer2020effect,tian2020investigation,ferretti2020quantifying,flaxman2020estimating,kissler2020projecting,walker2020impact}.
In contrast, in this work we focus on a simple class of policies for closing down and reopening the society throughout the course of a pandemic, thereby complementing the rich literature on covid modelling.

We consider infection dynamics of coronavirus in a population of size~$N$.
The population represents a community (city, state, or a country).
Initially all people are uninfected.
Then we add one (or several) infected individuals and follow the stochastic trajectories of viral spread.
The process advances in discrete time steps which represent days.
Individuals are in different states describing their status of being: susceptible (S), exposed/incubating (E), mildly ill/infectious (I), critically ill (C), and recovered/removed (R), see Figure~1a.
We assume that critical cases are hospitalized. 
The infection spreads whenever a susceptible person comes in contact with an infectious person.
In this case the infection is transmitted with probability~$p$ (see Figure~1b).
We denote the number of daily contacts per person by~$k_0$.


We assume that the community has a capacity~$c$ of hospital beds to treat the critical cases.
When left unregulated, the disease would surge through the community and 
exceed that capacity~$c$ (see Figure~1c).
A country can mitigate the spread of the disease by introducing various non-pharmaceutical interventions, such as
 enforcing social distancing or shutting down non-essential businesses.
 We model such interventions by decreasing the 
number~$k_0$ of daily contacts of an individual to a value~$\k<k_0$.
We call the regime when the interventions are put in place a \textit{lockdown}.

Here we study the key question of how quickly and how severely a community should lock down, and how patient it should be before reopening again.
To that end, we consider 
a simple class of policies characterized by three parameters~$\thr$,~$\k$,~$\wait$.
The parameter~$\thr$ is the number of critical cases that triggers the community to enter the lockdown.
It models the cautiousness or agility of the policy.
The parameter~$\k$ is the number of daily contacts per person in a lockdown.
It models the severity of the policy.
Finally, the parameter~$\wait$ is the number of days the community needs to spend with critical cases below the trigger threshold~$\thr$ before the lockdown is lifted.
It models the patience of the policy.
In other words, a policy~$P(\k,\thr,\wait)$
locks down to~$\k<k_0$ daily contacts per person
once the number of critically ill individuals exceeds a given trigger threshold~$\thr$, and it
reopens to~$k_0$ daily contacts once
the number of critically ill individuals remains under that threshold~$\thr$ for~$\wait$ consecutive days. 
We evaluate the performance of such policies with respect to several measures.
For instance, we consider the \textit{peak load}~$\peak$, which is the expected number of critical cases at its maximum, and
 the \textit{overflow probability}~$\pcollapse$, which is the probability that the peak load~$\peak$ exceeds the hospital bed capacity~$\capacity$ available to treat the critically ill cases.

We find that the best performing policies are those that quickly transition to a severe lockdown and that are patient about reopening. 
However, when either quick or severe action is not feasible, as is often the case, a country can compensate by pressing more along the other dimension.
This gives rise to a spectrum of possible policies for closing down and reopening the society.
At one end of this spectrum, a
\textit{moderate low-trigger} policy ($\quick$) imposes a gentle lockdown at the first sign of the onset of the disease.
At the other end, a \textit{severe high-trigger} policy ($\lazy$) remains open till the latest moment possible and then imposes a harsh lockdown.
We find that, though comparable in some regards, those two policies are very different in terms of
their long-term behavior and
in terms of their sensitivity to policies employed by neighboring countries.
Specifically, we argue that
the moderate low-trigger policies are preferable assuming that the countries are able to coordinate and that an efficient vaccine is not distributed soon, whereas the severe high-trigger policies are preferable otherwise.

\section{Model}

In order to describe the spread of COVID-19, we consider a stochastic, discrete-time, individual-based SIR-like model.

\sne{Disease progression within an individual}
Initially, a typical individual is susceptible~(S) and can contract the disease after contact with an infectious individual (see Figure~1a).
 Immediately upon contracting the disease, an individual becomes exposed/incubating (E) and does not yet spread it.
Later they become infectious (I) as they develop mild condition and then they either recover (R) or become critically ill (C). Critical individuals are hospitalized, isolated and they occupy part of the capacity~$\capacity$ of the health system. Eventually they either recover or die (R). We assume that recovered individuals acquire immunity.
Upon contracting the disease, each transition occurs after a number of days that is given by a corresponding random variable~$X_{E\to I}$,~$X_{I\to R}$,~$X_{I\to C}$,~$X_{C\to R}$.
For concreteness, we set the values based on the data on COVID-19:~\cite{guan2020clinical,zhou2020clinical,grasselli2020baseline,li2020early,rajgor2020many,verity2020estimates,ruan2020likelihood}
The incubation period is~$X_{E\to I}=2$ days and the individual recovers from a mild condition upon~$X_{I\to R}=10$ days.
During each of those 10 days, an individual might become critically ill with probability~$1\, \%$ 
(hence~$X_{I\to C}$ is exponentially distributed with parameter~$1\,\%$ and roughly 10 \% of cases become critical).
The critical cases recover (or die) after~$X_{C\to R}=10$ more days.

\sne{Disease spread through the population}
We consider a population of~$N$ individuals.
Each day, each individual comes in contact with~$k_0$ other individuals (see Figure~1b).
When a susceptible individual (S) meets an infectious individual (I), he or she contracts the disease with transmission probability~$p$ and becomes incubating (E).
We denote the number of critically ill individuals at day~$t$ by~$C(t)$ and use analogous notation for other conditions.
For concreteness~\cite{rhodes2012variability,ma2020critical,phua2020intensive} we consider the population of~$N=20\,000$ individuals, the health system capacity of 2.8 beds per 1\,000 individuals leading to~$\capacity=56$ beds in total,~$k_0=15$ daily interactions per person, and the transmission probability~$p=2\,\%$. All in all, this gives the epidemiological basic reproductive ratio~$R_0$ equal to roughly~$R_0=k\cdot p\cdot X_{M\to R} =2.9.$

\sne{Policies}
We consider a 3-parameter class of policies that a country can use to mitigate the spread of the disease.
The policies toggle between two regimes -- the default \textit{open} regime and the temporary \textit{lockdown} regime (see Figure~\ref{fig:basic}a).
The policy~$P(\thr,\k,\wait)$ can be efficiently described using three parameters~$\thr$,~$\k$,~$\wait$ that describe how soon and severely the policy locks down, and how soon it reopens:
\begin{enumerate}
\item Once the number~$C(t)$ of critically ill cases exceeds the trigger threshold~$\thr$, the policy locks down by reducing the number of daily contacts per person from~$k_0$ to~$\k$.
\item Once the number~$C(t)$ of critically ill cases remains under the trigger threshold~$\thr$ for~$\wait$ consecutive days, the policy reopens by resetting the number of daily contacts per person back to~$k_0$.
\end{enumerate}
The three parameters~$\thr$,~$\k$,~$\wait$ thus model three natural features of the policy: its ``cautiousness'', that is, how easily it is triggered into a lockdown;  the ``severity'' of its lockdowns; and its ``patience'' when reopening, respectively.
We remark that we chose the number of critically ill people~$C(t)$ rather than the number of infectious people~$I(t)$ since the latter is not so easily accessible to policymakers.

\sne{Performance of the policy}
In order to evaluate the performance of a policy, we study the following quantities:
\begin{enumerate}
\item The (expected) \textit{peak load}~$\peak$: That is, the expected number~$\peak=\E[\max \{ C(t) \mid t\ge 0 \}]$ of critically ill people at their maximum, over the duration of the disease. This represents the maximum demand on the health system (hospital beds).
\item The \textit{overflow probability}~$\pcollapse$: That is, the probability~$\pcollapse=\Pr[\peak>\capacity]$ of exceeding the available bed capacity~$\capacity$ at some point throughout the course of the disease.
\item The \textit{total load}~$\Call$: That is, the total cumulative number of critical cases, over the duration of the disease. This can be used to estimate the total number of deaths.
\item The total (expected) \textit{duration}~$\dur$ of the lockdown: That is, the total expected number of days spent in the lockdown regime until the disease is eradicated.
\end{enumerate}
In all cases, the lower the quantity the better the policy. Hence we can think of all the quantities as costs of the policy.
We note that the first three quantities can be viewed as costs related to health of the population.
The fourth quantity can be viewed as an economic cost of imposing a lockdown of a given severity, since a lockdown reduces the economic activity of a community.
 
\section{Results}

We evaluate the performance of the above defined policies -- first for a single country, later for two neighboring countries.
Recall that a policy~$P(\thr,\k,\wait)$ is given by three parameters:
The threshold number of critical cases~$\thr$ that triggers the policy to toggle to lockdown and back (``trigger value'');
the number~$\k$ of daily contacts per person during a lockdown (``severity''); and
the number~$\wait$ of days required to remain below the trigger threshold~$\thr$ before the society can reopen (``patience''),
see Figure~\ref{fig:basic}a.

\sne{Four example policies} To illustrate the differences in performance of various policies, we first consider four specific policies that all share the patience parameter~$\wait=10$ days and that differ in the trigger value~$\thr$ and in the severity~$\k$ (see Figure~\ref{fig:basic}b).
Specifically, in terms of the trigger value~$\thr$, we distinguish
\textit{low-trigger} policies ($\thrlowname=\thrlow$) from \textit{high-trigger} policies~$(\thrhighname=\thrhigh)$.
Similarly, in terms of the number~$\k$ of daily contacts in a lockdown, we distinguish
\textit{severe} policies~$(\klowname=\klow)$ from \textit{moderate} policies~$(\khighname=\khigh)$.
All in all, this yields~$2\times 2=4$ combinations 
$\best$ (severe, low-trigger),
$\lazy$ (severe, high-trigger),
$\quick$ (moderate, low-trigger), and
$\worst$ (moderate, high-trigger).
We observe that the policies substantially alter how the number of infected and critical cases evolves in time (see Figure~\ref{fig:basic}c).


To explain the difference, it is instructive to think in terms of the epidemiological~$R_0$ that determines whether the number of infected individuals in a population is quickly surging ($R_0> 1$), disappearing ($R_0< 1$) or changing slowly ($R_0\approx 1$).
Note that~$R_0$ is not constant in time -- it crucially depends on the current number~$k$ of daily contacts ($R_0$ decreases as~$k$ decreases)
and also on the percentage~$x$ of immune individuals ($R_0$ decreases as~$x$ increases).
In the open society ($k$ large) and with no immune individuals ($x=0$) we have~$R_0> 1$,
hence the disease initially spreads quickly.

\snt{Severe policies}
Under the two severe policies, the number of cases in time follows the familiar spikes:
Each lockdown is so harsh that as long as it is in place, we have~$R_0< 1$ even when~$x=0$. 
Therefore, a few days upon imposing the lockdown (the incubation period) the infected cases rapidly drop, then the critical cases drop too and the disease can possibly get eradicated in some communities.
This happens over a short period of time and only a few people acquire immunity ($x\approx 0$).
When the trigger~$\thr=\thrhighname$ is high, the patience of~$\wait=10$ days is insufficient to eradicate the disease completely, the lockdown is lifted too early, a subsequent spike of similar shape is likely, and the whole cycle repeats several times.
When the trigger~$\thr=\thrlowname$ is low, waiting for~$\wait=10$ days will typically suffice to eradicate the disease completely and no subsequent spikes occur.

 \snt{Moderate policies}
Under the two moderate policies, the typical stochastic trajectories are different.
The lockdown is so gentle that when~$x=0$, we have~$R_0\approx 1$.
Hence upon imposing a moderate lockdown, the number of ill individuals becomes roughly constant in time.
But as time goes by and the individuals progress through the disease stages and acquire immunity ($x>0$), the value of~$R_0$ decreases and the disease starts to die out ($R_0< 1$).
Therefore, compared to the severe policies, the first peak is substantially broader.
(However, the taller it is, the less apparent this distinction is.)
Crucially, if the lockdown is lifted too soon and another outbreak occurs later, once the same moderate lockdown is imposed again,
the immune subpopulation ($x>0$) causes the disease to die out right away -- in fact, it dies away faster and faster in every subsequent lockdown.
Therefore while the first outbreak might require a long lockdown phase, all subsequent outbreaks are dealt with promptly.
In a sense, when the moderate lockdown is lifted for the first time, the population had already acquired herd immunity level for the interaction rate~$k$ of the moderate lockdown.
This means that when the lockdown is put in place, the disease does not spread. However, once the lockdown is lifted and individuals start to interact more frequently, the disease could start spreading again.
For our parameters, when the trigger~$\thr=\thrhighname$ is high, the lockdown is not strong enough and the bed capacity is (slightly) exceeded within the first peak.
When the trigger~$\thr=\thrlowname$ is low, the critical cases stay safely below the bed capacity.
In both cases, the patience~$\wait=10$ days is insufficient to completely eradicate the disease and subsequent peaks occur -- all substantially smaller than the first one.

\sne{How parameters affect the performance}
In order to understand the role of the three parameters (trigger value~$\thr$, severity~$\k$, patience~$\wait$) on each of the four performance measures (peak size~$\peak$, overflow probability~$\pcollapse$, total number of critical cases~$\Call$, lockdown duration~$\dur$), we run exhaustive computer experiments.

In Figure~\ref{fig:trigger}, each row shows a different performance measure ($y$-axis) as a function of the number~$\k$ of daily contacts in a lockdown ($x$-axis).
The red dotted vertical line marks the number~$\k^\star$ of daily contacts that corresponds to~$R_0=1$ (when there are no immune individuals,~$x=0$).
Within each row, the left panel shows policies that have low patience ($\wait=7$ days) and the right panel shows policies that have high patience ($\wait=70$ days).
Within each panel, the blue curve shows the low-trigger policies ($\thrlowname=\thrlow$) and the green curve shows the high-trigger policies ($\thrhighname=\thrhigh$).


Recall that for each performance measure, the lower the value, the better the performance.
The effect of the trigger value can be seen by comparing the blue and the green curve:
Since the blue curve is typically lower, low-trigger policies are generally better than high-trigger policies.
The effect of the lockdown severity can be seen by looking at whether the curves go up or down:
Since they typically go up, severe policies are generally better than moderate policies.
The effect of the patience can be seen by comparing the left and the right panel:
Since the curves in the right panel are typically lower, patient policies are generally better than impatient policies.

There are two exceptions to those rules, both concerning the lockdown duration~$\dur$: First, we observe that when~$\k$ becomes too large, the duration~$D$ decreases. This is due to the fact that the lockdown becomes too weak and the disease quickly sweeps through the whole population. Second, we observe that when the patience is low and the lockdown is moderate, decreasing the trigger value~$\thr$ actually leads to more time spent in the lockdown.

\sne{Key parameters for different regimes}
Next, for each performance measure, we characterize which of the three parameters~$\thr$,~$\k$,~$\wait$ are key to substantially improving the performance and which of them are marginal. 

Let~$\k^\star$ be the number of daily contacts that corresponds to~$R_0=1$ when there are no immune individuals ($x=0$).
For our parameters, we have~$\k^\star\doteq 5.3~$.

\snt{Peak size~$\peak$} 
We observe that (see Figure~\ref{fig:trigger}a):
\begin{enumerate}
\item[$\thr$:] The low-trigger policies (blue curve) are consistently better than the high-trigger policies (green curve).
\item[$\k$:] For both trigger values,~$\peak$ is roughly constant as long as~$k<k^\star$ but then it increases rapidly when~$k>k^\star$.
\item[$\wait$:] The left and the right panel are comparable.
\end{enumerate}
Hence the important insight is to have the severity below a threshold ($\k<\k^\star$) and to have the trigger~$\thr$ low.
The effects of the patience~$\wait$ and the severity~$\k$ (given that~$\k<\k^\star$) are marginal.

The intuitive explanation is that 
when~$\k>\k^\star$, the lockdown is so weak that the disease still continues to spread, even when the lockdown is put in place.
Hence having~$\k<\k^\star$ is key.
 On the other hand, given that~$\k<\k^\star$, the actual value of~$\k$ is not that important:
When a lockdown is put in place, the peak size (and the moment when the peak occurs) have already been essentially determined, since most of the critical cases at the peak are due to individuals who have already been infected when the lockdown was put in place. 
Similarly, the patience~$\wait$ is not that important as it affects the number of peaks rather than their size.
On the other hand, the trigger value~$\thr$ is key:
The nature of the exponential growth and the inherent delay due to incubation and non-critical infection translate the difference in trigger value~$\thr$ to a difference in peak size~$\peak$.

\snt{Overflow probability~$\pcollapse$}
We observe that (see Figure~\ref{fig:trigger}b):
\begin{enumerate}
\item[$\thr$:] The low-trigger policies (blue curve) are consistently better than the high-trigger policies (green curve).
\item[$\k$:] For low-trigger policies,~$\pcollapse$ exhibits a threshold behavior with respect to~$\k$.
For high-trigger policies,~$\pcollapse$ increases when~$\k<\k^\star$ and increases rapidly when~$\k>\k^\star$.
\item[$\wait$:] For high-trigger policies (green curve), increasing the patience~$\wait$ decreases~$\pcollapse$ (when~$\k<\k^\star$).
\end{enumerate}
Hence the important insight is, again, to have the severity below a threshold ($\k$ below~$\k^\star$ or just very slightly above) and to have the trigger~$\thr$ low.
When the trigger~$\thr$ is high, both decreasing~$\k$ and increasing~$\wait$ help, but even the combined effect is negligible compared to the effect of having~$\thr$ low.

The intuitive explanation is that
in large populations, most stochastic trajectories are qualitatively similar.
Hence even though the peak size is a random variable, it is narrowly concentrated around its average value.
Thus whenever the average peak size~$\peak$ slightly exceeds the available bed capacity~$\capacity$,
the overflow probability is almost 1.
And, vice versa, whenever the average peak size~$\peak$ is slightly lower than the available bed capacity~$\capacity$,
the overflow probability is almost 0.
In a sense, the overflow probability exaggerates the difference between~$\peak$ and~$\capacity$.

\snt{Total critical cases~$\Call$}
We observe that (see Figure~\ref{fig:trigger}c):
\begin{enumerate}
\item[$\thr$:] When impatient (left panel), the low-trigger policies (blue curve) are consistently better than the high-trigger policies (green curve).
\item[$\k$:]  For impatient high-trigger policies (green curve in the left panel),~$\Call$ is roughly constant unless~$\k$ is small. 
\item[$\wait$:] Increasing the patience~$\wait$ decreases~$\Call$.
\end{enumerate}
Hence the important insight is to have the patience~$\wait$ high.
In order to achieve comparable results with low patience, one has to have a low trigger~$\thr$ and a very severe lockdown.

The intuitive explanation is that
high patience~$\wait$ is key because it decreases the chance of a premature reopening and thereby reduces the number of times a lockdown has to be put in place.
With a sufficiently high patience, the disease gets eradicated upon completing the first peak.
With low patience~$\wait$ (and thus many peaks), the only way to avoid many critical cases is to make sure each peak is small. This requires a low trigger value~$\thr$ and severe lockdown~$\k$ every time the trigger value is reached.
As a final remark, note that when impatient and high-trigger, the lockdowns have to be extremely severe to help even a little: This is the only way to get at least some hope that the disease gets eradicated by the short time the lockdown will be lifted.

\snt{Total lockdown duration~$\dur$}
Since a severe lock down is very different from a moderate lockdown, here we focus our comparison on only those lockdowns that have the same severity.
We observe that (see Figure~\ref{fig:trigger}d):
\begin{enumerate}
\item[$\thr$:] When impatient (left panel), low-trigger policies (blue curve) are better assuming~$\k$ is small, otherwise high-trigger policies (green curve) are better. When patient (right panel), low- and high- trigger policies are comparable.
\item[$\wait$:] Increasing the patience~$\wait$ decreases~$\dur$. 
\end{enumerate}
Hence, for severe lockdowns ($\k< \k^\star$), the important insight is, again, to have the patience~$\wait$ high.
In those cases, the effect of the trigger value is marginal.

The intuitive explanation for severe lockdowns of fixed severity~$\k< \k^\star$ is that the key to minimizing the total lockdown duration is to minimize the probability~$q$ of a subsequent outbreak (that would occur if the lockdown were lifted too early).
Since in a severe lockdown, the numbers of infected and incubating individuals decay exponentially ($\k<\k^\star$), there are two ways to decrease~$q$ (and thereby~$\dur$) by a constant factor:
Either to increase the patience by a constant number of days, or
to decrease the trigger value~$\thr$ by a constant factor.
The former is less costly, hence having high patience is the single most important aspect.

\snt{Summary}
Here we summarize three findings from the above paragraphs.
First, any successful policy must impose lockdowns such that restrict the number~$\k$ of daily contacts under the threshold~$k^\star$.
Second, in terms of minimizing either the peak size~$\peak$ or the overflow probability~$\pcollapse$, it is crucial to employ a policy with low trigger threshold~$\thr$.
Third, in terms of minimizing either the total number~$\Call$ of critical cases or the lockdown duration~$\dur$, it is crucial to employ policies with high patience~$d$. The precise optimal value of patience depends on the severity of the lockdown and on the trigger threshold~$\thr$.
Generally speaking, moderate policies ($\khighname=\khigh$) benefit from patience as large as~$d=56$ days (8 weeks).
In contrast, for severe policies~$(\klowname=\klow$), a patience of 8 weeks gives comparable results to a patience of 4 weeks, see Figure~\ref{fig:damage}.


\sne{Two countries}
Some interaction among communities is inevitable. While the inter-community interaction can be limited by closing borders between countries and imposing  quarantine upon entry, it can not be completely disregarded.
We study how the policies perform in the environment where different communities might employ different policies.

To model this, we consider two communities Country 1 and Country 2 that experience the onset of the disease at the same time.
Occasionally, individuals from different countries meet. Namely, we assume that for each individual, a small portion~$\kacross\in(0,1)$ of their interactions are with individuals in the other country.

We consider two very different policies and study how their performance depends on the policy employed by the neighboring country and on the interaction rate~$\kacross$ between the two countries (see Figure~\ref{fig:2-countries}). 

Specifically, we consider s moderate low-trigger policy~$\quick=P(\thr=\thrlow,\k=\khigh,d=54)$ and a severe high-trigger policy~$\lazy=P(\thr=\thrhigh, \k=\klow, d=27)$.
The parameters~$\thr$ and~$\k$ are chosen such that both policies have the same probability~$10\ \%$ of exceeding the US hospital bed capacity within their first peak.
The parameter~$\wait$ is chosen such that, on average,
$90\ \%$ of the critical cases occur within the first peak.
In other words, the policy reopens when it expects that 90 \% of all the critical cases have already been hospitalized.


When both countries employ the same policy, the performance is essentially the same as for a larger country employing that policy.
Also, in the limit~$\kacross\to 0$ the two countries do not interact and the performance of a country is independent of the policy of the neighbor (this remains true for~$q<10^{-5}$).
However, when~$\kacross$ is non-negligible, one country uses~$\quick$ (``$\quick$-country'') and the other one uses~$\lazy$ (``$\lazy$-country''), the two policies clash. Specifically, we make the following observations about the overflow probability~$\pcollapse$ and the average peak size~$\peak$  (see Figure~\ref{fig:2-countries}):

\begin{enumerate}
\item For~$\kacross$ small, the green curve goes up: Note that the infectious subpopulation of the neighboring~$\quick$-country is non-negligible for an extended period of time (at least throughout the first broad peak). Thus, as~$\kacross$ increases, the individuals in the~$\lazy$-country get repeatedly infected due to interactions with the~$\quick$-country.
Most such new infections cause a new spike for the~$\lazy$-country. Each such spike might exceed the previously largest peak and/or the available capacity.
(Moreover, each such spike leads to new critical cases and it has to be contained by another lockdown phase so it is costly in terms of~$\Call$ and~$\dur$ too.)
 This effect is visible for interaction rates as small as~$\kacross=10^{-4}$.
 
\item For~$\kacross$ large, the blue and green curve go down:
Two countries employing different policies typically reach their peaks at a different point in time. Thus, when one country is peaking, the other country likely has fewer infectious individuals and an interaction with that country will help alleviate the size of the peak in the first country. This, in turn, decreases the maximum peak load~$\peak$ and the overflow probability~$\pcollapse$.
This effect is visible roughly for~$\kacross>10^{-3}$ (blue curve), resp.~$\kacross>10^{-1}$ (green curve).

\item The yellow and the red curves are roughly constant, except that the yellow curve goes down in terms of the overflow probability when~$\kacross$ is large:
When two neighboring countries employ the same policy, the extra occasional mixing due to interacting individuals makes both countries behave in a slightly more average way.
This does not change the expected size of the peak, hence~$\peak$ is constant, regardless of the interaction rate~$\kacross$.
However, this process of ``averaging out'' does make the extreme behavior, such as overflowing of the available hospital capacity, somewhat less likely.
This effect of diminished overflow probability is observed for two~$\quick$ countries when~$\kacross>10^{-2}$.
\end{enumerate}

This leads to an interesting phenomenon resembling a social dilemma.~\cite{nowak2006five,sigmund2016calculus}
Consider two countries with a non-negligible interaction rate~$\kacross>10^{-3}$.
First, if either of the two countries primarily cares about keeping the overflow probability low then that country would employ the policy~$\quick$ rather than the policy~$\lazy$, no matter which policy the other country is using (indeed blue is below red and yellow is below green).
Second, given that one country uses the policy~$\quick$, the other country will use the policy~$\quick$ too (yellow is below green).
In terms of the overflow probability alone, this is an acceptable outcome (red and yellow are comparable) but in terms of the total number of critical cases~$\Call$, this is undesirable:
By employing the policy~$\quick$ rather than~$\lazy$, both countries increase the total number of their critical cases (and thus, possibly, deaths) by an order of magnitude.

\section{Discussion}
Motivated by the COVID-19, we studied a simple stochastic model of a disease progression
in a population of interacting individuals.
We focused on a 3-parameter family of policies that can be used to mitigate the disease spread and evaluated the performance of those policies with respect to several measures, such as the number of critical cases at its maximum or the probability that this number exceeds the available hospital bed capacity.
The three parameters describing the policies correspond to
 the agility when closing down, 
 the severity of the lockdown, and 
 the patience when reopening, 
We identified which parameters are important in which regime and explained why some policies are performing better than others.

We highlight two different types of realistic policies, called \textit{moderate low-trigger} ($\quick$) and \textit{severe high-trigger} ($\lazy$).
With both policies, the probability~$\pcollapse$ of ever exceeding the available hospital beds is kept below a specified threshold (here arbitrarily set to 10 \%), but the two policies are very different:
The~$\lazy$ policy is characterized by imposing a harsh, short lockdown (``severe'') at the last moment possible (``high-trigger''), whereas
the~$\quick$ policy is proactive and imposes gentle, longer lockdowns (``moderate'') at the first signal of an approaching outbreak (``low-trigger'').

Due to the above differences, both policies have their advantages and disadvantages.
The~$\lazy$ policy minimizes the total number of critically ill cases and the total amount of time spent in lockdown.
However, the lockdowns are severe and the society is more susceptible to any subsequent outbreaks.
To avoid such recurring outbreaks, the authorities must be patient when reopening and any neighboring countries must coordinate when releasing their measures.
The~$\quick$ policy maintains its moderate lockdown for substantially longer but its performance is
substantially more robust with respect to how soon the society reopens and with respect to
what policies are employed by the neighboring countries.
Moreover, upon completing the first lockdown phase, 
 all subsequent lockdowns (if any) are shorter and involve substantially fewer critical cases than the first phase.
Thus the~$\quick$ policy can be seen as minimizing the long-term risks under the pessimistic scenario that an ultimate long-term solution (such as a majority of the population being vaccinated) is not achieved any time soon.
On the other hand, the~$\lazy$ policy is optimistic about the future and optimizes the short-term performance.

Our setup is intentionally simplified in many regards such as the disease progression within an individual, the disease spread throughout the population, and the class of policies we consider for closing down and reopening.
We highlight four possible extensions worth pursuing in subsequent work:
First, regarding the disease progression within an individual, one can
distinguish more types of individuals (e.g. those who require only hospital beds and those who moreover require ventilation) and lift the assumption that the individuals acquire lifelong immunity.
Second, regarding the disease spread in a population, one can
consider a population structure, e.g. described by a graph whose edge weights determine the daily pairwise transmission probabilities.
This would allow one to investigate the effects of localized interventions such as contact tracing.
Third, one can
consider more complicated policies, e.g. policies that allow for a gradual reopening or policies that, when deciding whether and how much to reopen, take into account additional information, such as the situation in the neighboring countries and/or the outcomes of testing done earlier.
Fourth, on top of considering the health viewpoint, one can incorporate the economic viewpoint by introducing an appropriate notion of economic cost of a lockdown of varying degree of severity.


\bibliographystyle{vancouver}
\bibliography{refs}


\pagebreak 
\begin{figure}[h]
\includegraphics[scale=0.5]{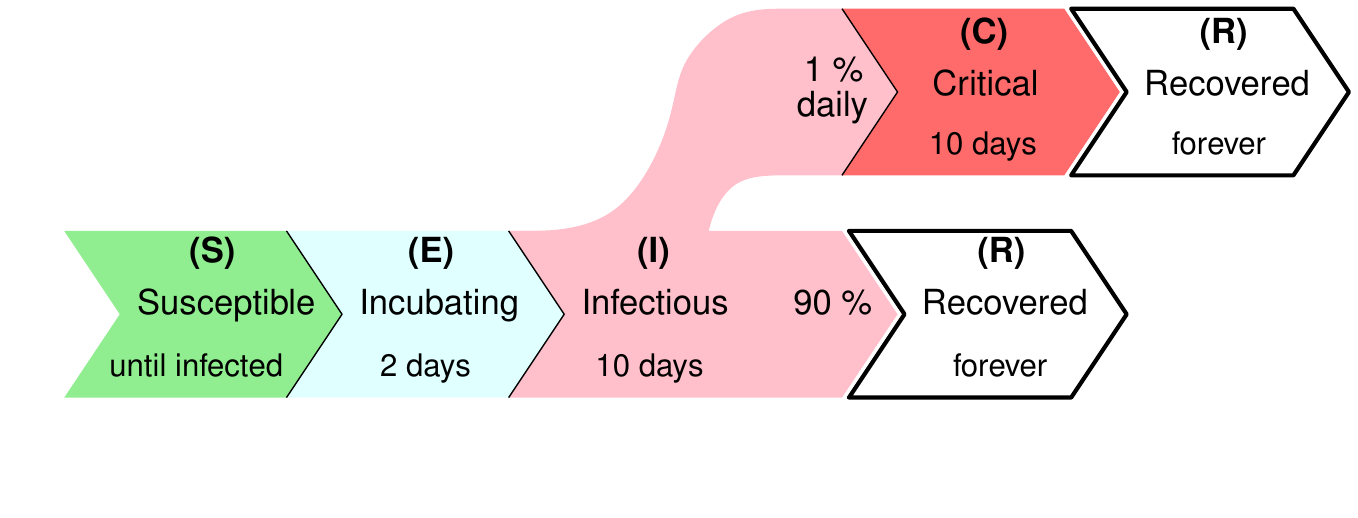} 
\includegraphics[scale=0.5]{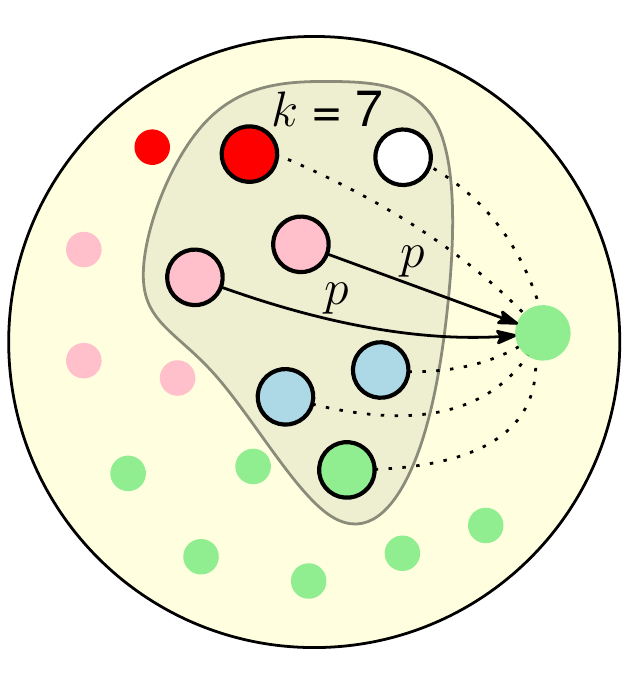} 
  \includegraphics[scale=0.5]{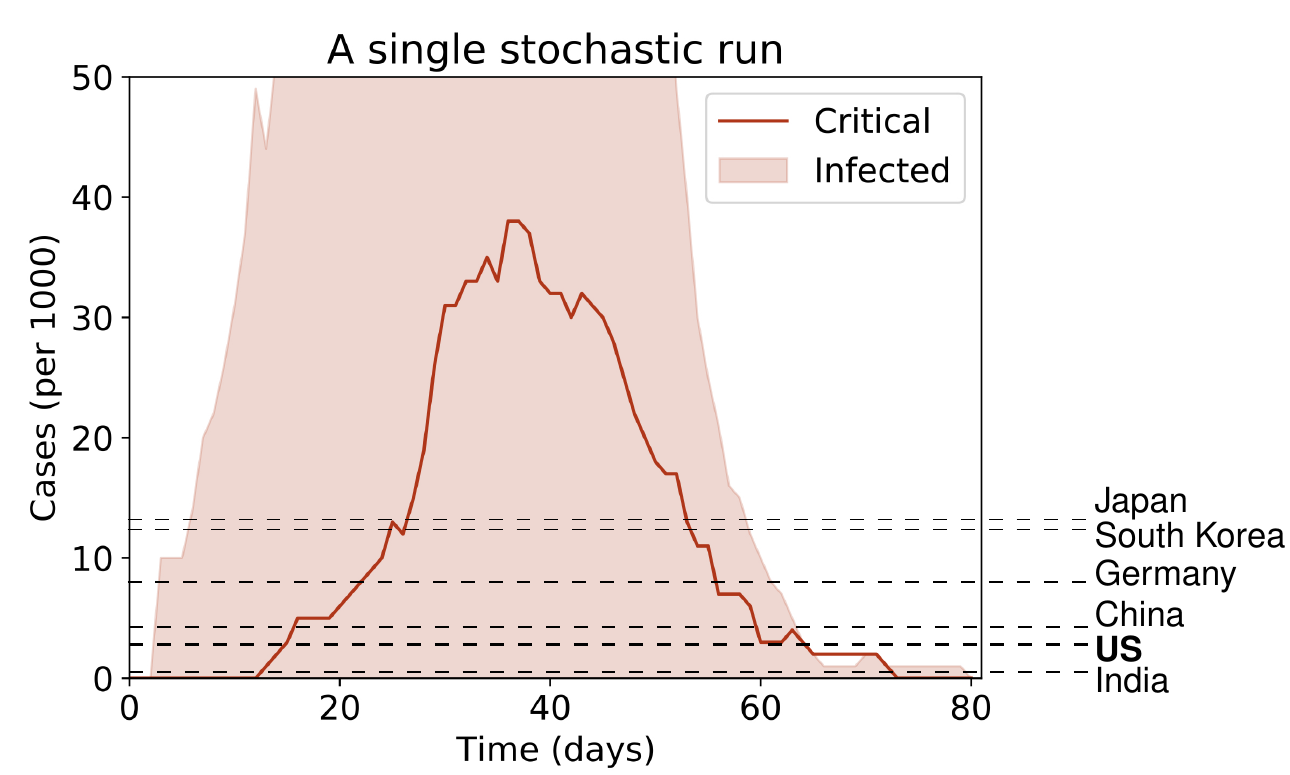}
\caption{\footnotesize{
\textbf{The disease spread without an intervention.}
\textbf{a,}~Upon contracting the disease, the individual incubates for two days (E) and then becomes infectious (I) as they develop mild condition. If a critical condition develops (C), the individual is hospitalized and isolated. We assume that all surviving individuals (R) acquire immunity.
\textbf{b,}~A population of~$N$ individuals. Each day, an individual meets~$k$ other individuals. During a single meeting with an infectious person, a susceptible individual contracts a disease with a transmission probability~$p$.
\textbf{c,}~Without intervention, the disease surges through the community and the critical cases (curve) at its peak~$\peak$ exceed the available hospital bed capacity~$\capacity$ (dashed lines~\cite{rhodes2012variability,ma2020critical}). Here~$N=1000$,~$k=15$,~$p=2\ \%$, thus the epidemiological~$R_0$ is roughly~$R_0\doteq 2.9$.  
 }}
\end{figure}

\pagebreak 
\begin{figure}[h]
\includegraphics[scale=0.8]{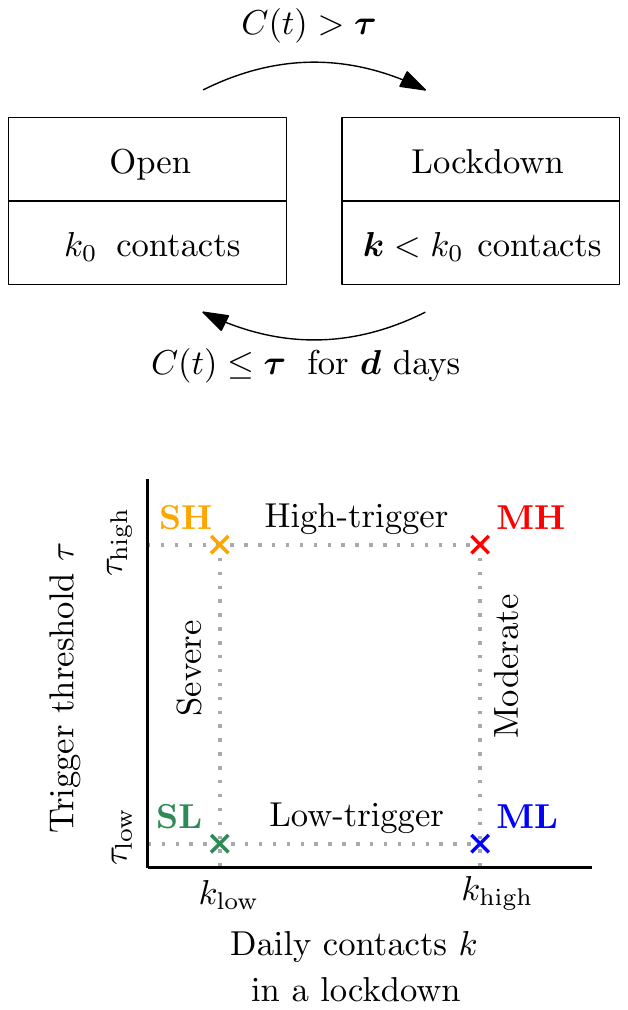}
\includegraphics[scale=0.8]{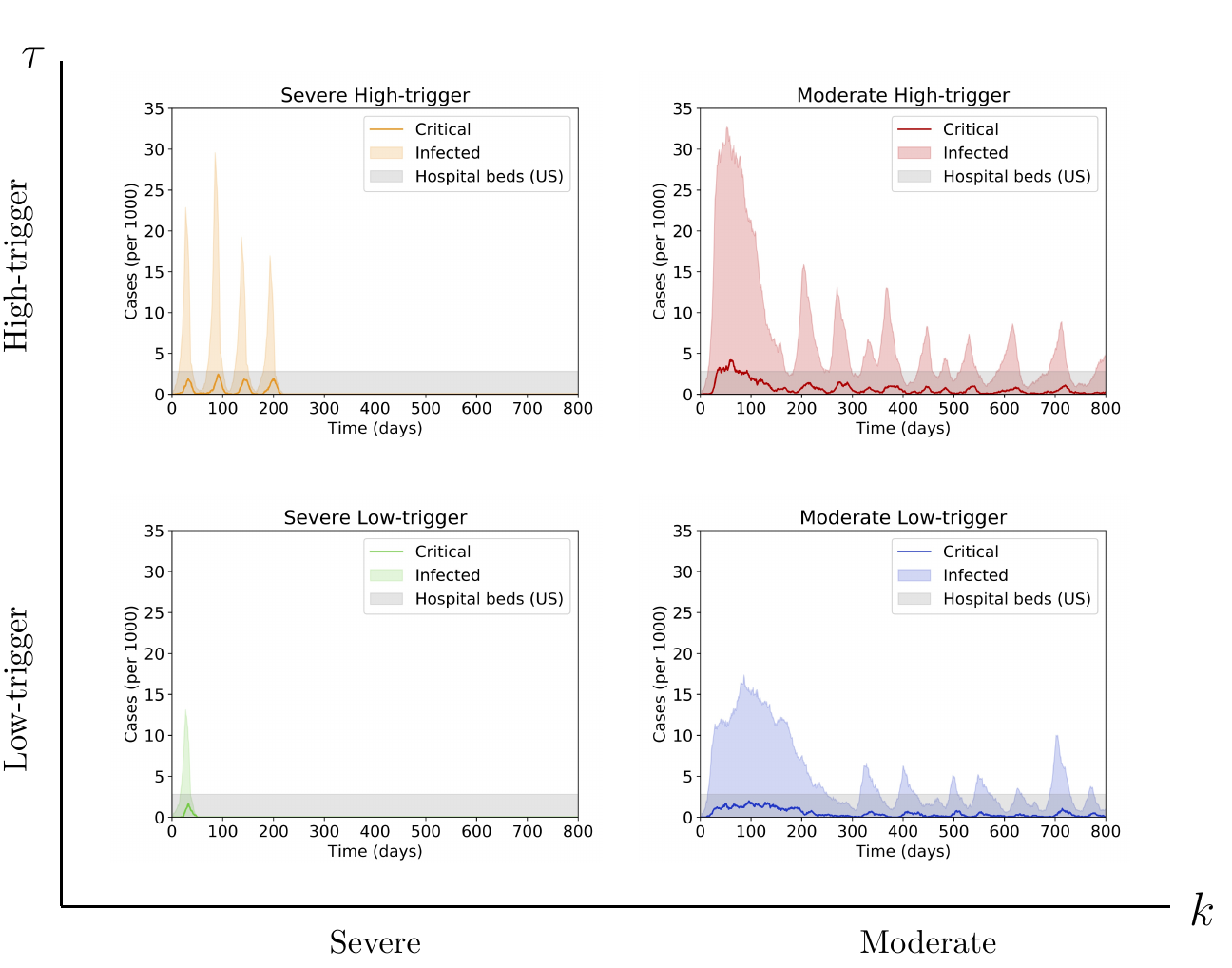}
\caption{\footnotesize{
\textbf{Basic policies.}
\textbf{a,} Under a policy~$P(\thr,\k,\wait)$, the country locks down to~$\k<k_0$ daily contacts whenever the number~$C(t)$ of critical cases exceeds a trigger threshold~$\thr$. It reopens (to~$k_0$ daily contacts) once the number of critical cases stays below~$\thr$ for~$\wait$ consecutive days.
\textbf{b,} Representative runs of four different policies given by a combination of a trigger threshold ($\thrlowname=\thrlow$,~$\thrhighname=\thrhigh$) and a lockdown severity ($\klowname=\klow$,~$\khighname=\khigh$), and common patience~$\wait=10$ days.
While with~$\lazy=P(\thrhighname,\klowname,10)$ (top left) all peaks are similar in shape, with~$\quick=P(\thrlowname,\khighname,10)$ (bottom right) all subsequent peaks are much smaller than the first one. With~$\worst=P(\thrhighname,\khighname,10)$ (top right), the capacity is exceeded and with~$\best=P(\thrlowname,\klowname,10)$ (bottom left), the disease is quickly eradicated.
}}
\label{fig:basic}
\end{figure}

\pagebreak 
\begin{figure}[h]
\includegraphics[scale=0.6]{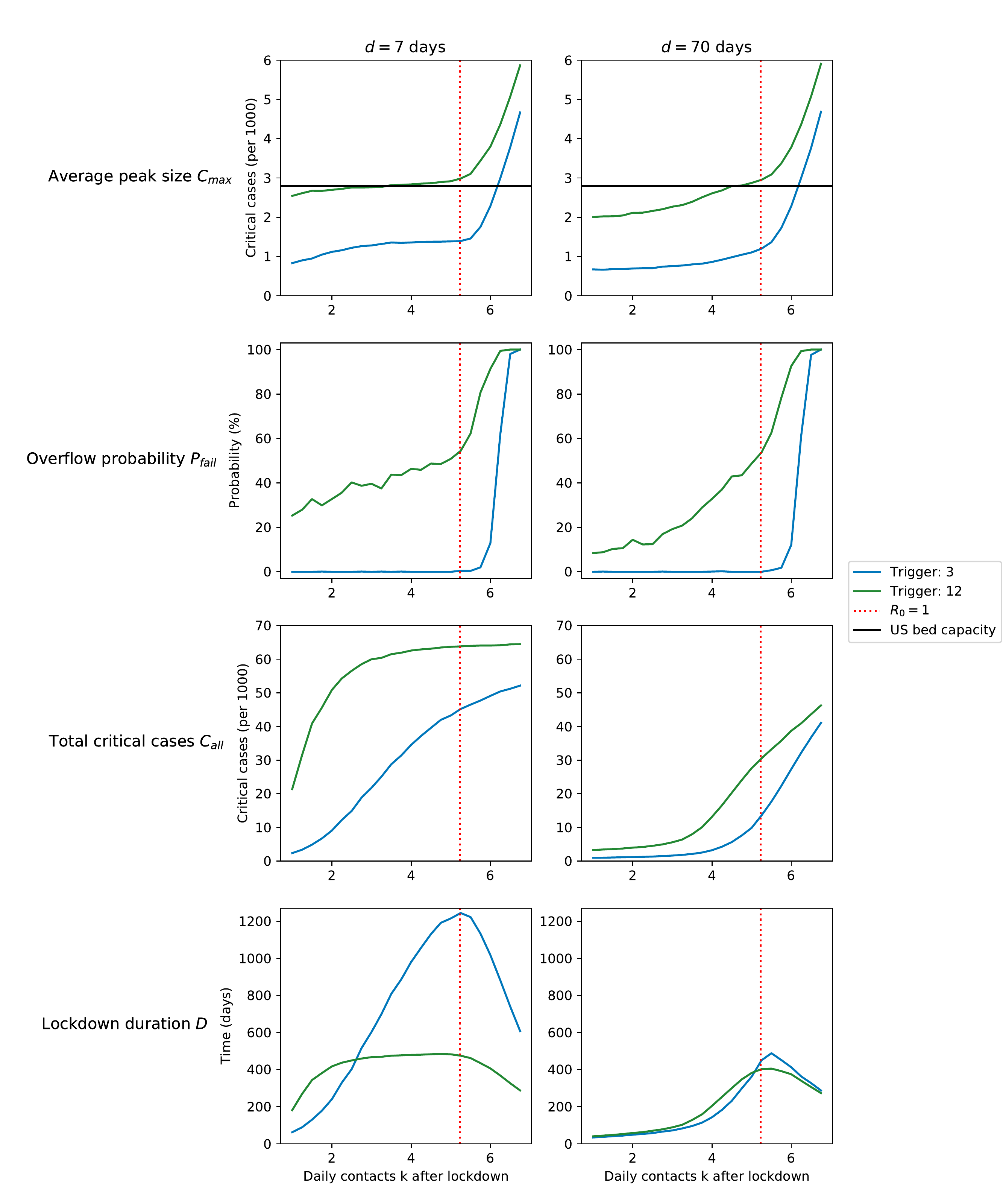}
\caption{\footnotesize{
\textbf{Performance of low-trigger and high-trigger policies.}
In each panel, we vary the number~$\k$ of daily contacts ($x$-axis) and consider the performance (cost) of the low-trigger policies ($\thrlowname=\thrlow$, blue) and of the high-trigger policies ($\thrhighname=\thrhigh$, green), when the patience parameter is low ($\wait=\waitlow$ days, left column) and high ($\wait=\waithigh$, right column).
The dotted red line shows the number~$\k^\star$ of daily contacts that corresponds to the epidemiological~$R_0$ equal to~$1$ (when no individuals have yet recovered).
Generally speaking, it is beneficial to have the trigger value~$\thr$ low (blue curves are below green ones), to impose severe rather than moderate lockdown (all curves go up), and to be patient (the curves in the right panels are lower).
For~$\peak$ and~$\pcollapse$, the key is to have the trigger value~$\thr$ low.
For~$\Call$ and~$\dur$, the key is to have the patience~$\wait$ high.
}}
\label{fig:trigger}
\end{figure}

\pagebreak 
\begin{figure}[h]
\includegraphics[scale=0.9]{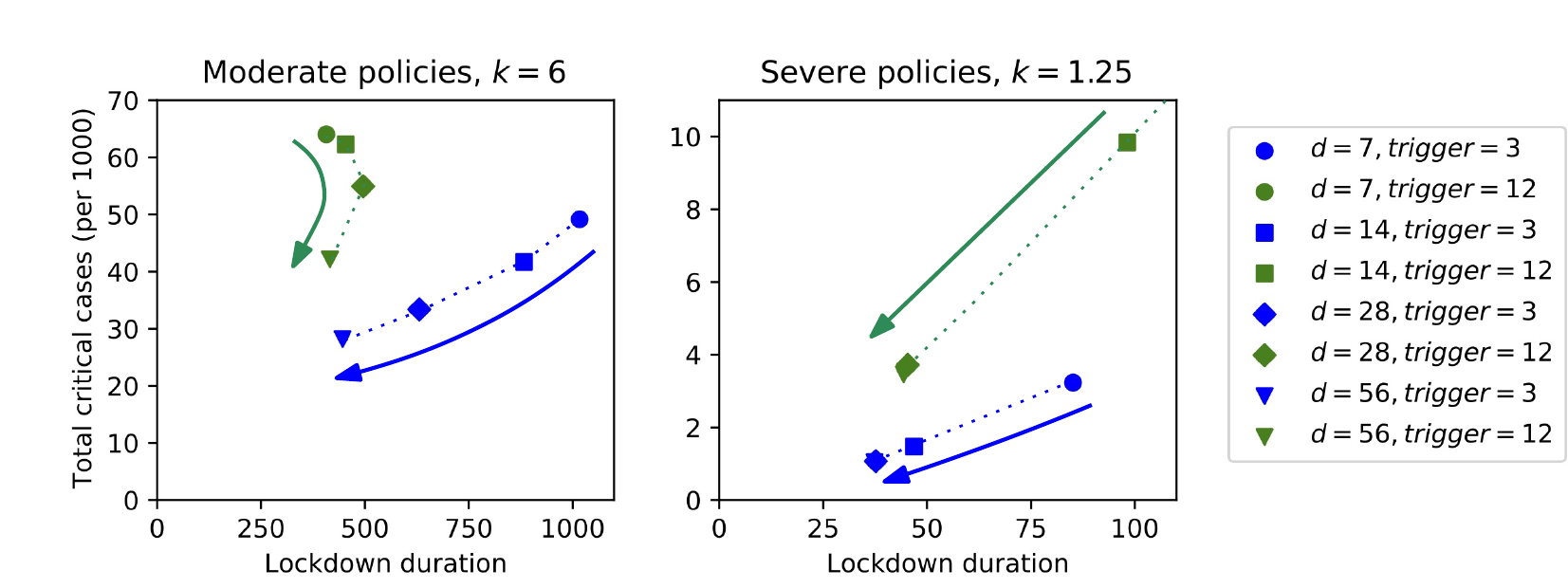}

\caption{\footnotesize{
\textbf{Role of patience in~$\Call$ and~$\dur$.}
The patience parameter~$d$ is key in determining the total health damage (total number~$\Call$ of critical cases per 1000) and the total economic damage (total duration~$\dur$ of the lockdown, in days).
The two panels are incomparable (due to different lockdown severity).
The low-trigger policies (blue) and the high-trigger policies (green). The patience increases along the arrow.
Features: High patience always helps in~$\Call$. High patience helps in~$\dur$, unless high-trigger and moderate (then it does not matter).
In the right panel, the diamond and the triangle marks coincide.
The high-trigger policy with low patience ($d=7$, green disk) is off the chard: It would yield~$D\doteq 270$ days and~$\Call\doteq 30$ cases per 1000.
}}
\label{fig:damage}
\end{figure}

\pagebreak 
\begin{figure}[h]
\includegraphics[scale=0.63]{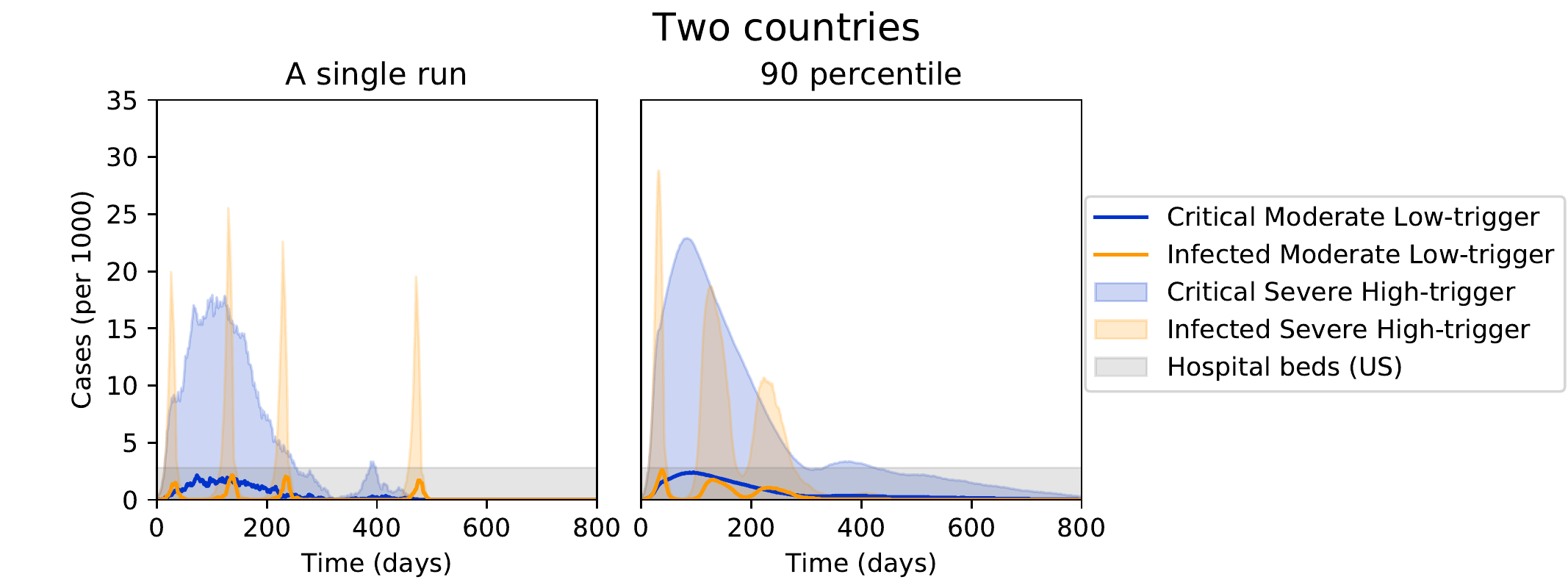}
\includegraphics[scale=0.63]{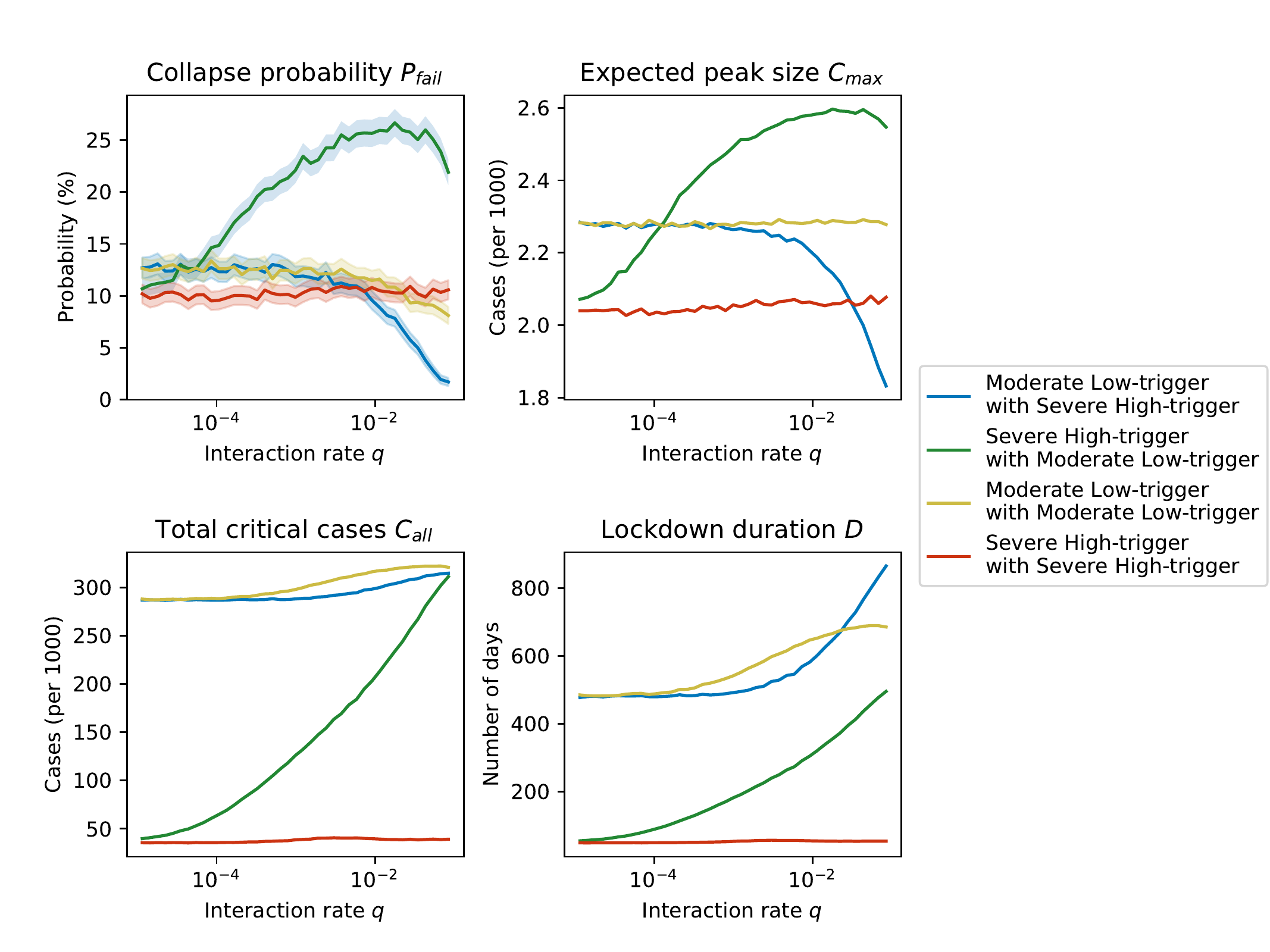}
\caption{\footnotesize{
\textbf{Two countries.}
\textbf{a,} When two countries interact with a positive rate~$\kacross$ they might reinfect each other.
Here~$\kacross=5\cdot 10^{-4}$, and the countries employ different policies:~$\quick$ (blue) and~$\lazy$ (yellow).
\textbf{b,} A~$90\ \%$ percentile: On any given day, 90 \% of the runs are below the respective curves.
\textbf{c-f,} The performance of a~$\quick$ policy against a~$\lazy$ policy (blue),~$\lazy$ vs. $\quick$ (green),~$\quick$ vs. $\quick$ (yellow) and~$\lazy$ vs. $\lazy$ (red), averaged over~$10^4$ runs. We vary the interaction rate~$\kacross$ on a log-scale and measure:
\textbf{c,} the collapse probability~$\pcollapse$  (95 \% confidence intervals are shaded);
\textbf{d,} the expected peak size~$\peak$;
\textbf{e,} the total number~$\Call$ of critical cases; and
\textbf{f,} the lockdown duration~$\dur$.
A country employing~$\lazy$ does great when its neighbor employs~$\lazy$ (red) but bad when the neighbor employs~$\quick$ (green).
A country employing~$\quick$ does comparably well, regardless of whether the neighbor employs~$\quick$ (yellow) or~$\lazy$ (blue).
}}
\label{fig:2-countries}
\end{figure}

\end{document}

%% file: macros.tex
\usepackage[usenames,dvipsnames]{xcolor}

\newcommand{\hide}[1]{}

\def\sne#1{\medskip\noindent\textbf{#1.}\ }
\def\snt#1{\medskip\noindent\textit{#1.}\ }

\def\peak{X_{\operatorname{peak}}}

\def\E{\mathbb{E}}

\def\thr{\tau}
\def\wait{d}
\def\peak{C_{\textrm{max}}}
\def\Call{C_{\textrm{all}}}
\def\capacity{c}
\def\pcollapse{p_{\text{fail}}}
\def\dur{D}
\def\quick{P_\mathit{ML}} 
\def\lazy{P_\mathit{SH}} 
\def\best{P_\mathit{SL}} 
\def\worst{P_\mathit{MH}} 
\def\kacross{q}

\def\k{k}
\def\low{\operatorname{low}}
\def\high{\operatorname{high}}

\def\thrlowname{\thr_{\low}}
\def\thrhighname{\thr_{\high}}
\def\klowname{\k_{\low}}
\def\khighname{\k_{\high}}

\def\thrlow{3}
\def\thrhigh{12}
\def\khigh{6}
\def\klow{1.25}
\def\waitlow{7}
\def\waithigh{70}

\usepackage{tikz}